\documentclass[10pt,twocolumn,letterpaper]{article}

\usepackage[pagenumbers]{wacv} % To force page numbers, e.g. for an arXiv version

\usepackage{color}
\usepackage{colortbl}
\usepackage{tabularx}
\usepackage{nicematrix}
\usepackage{balance}
\usepackage{listings} % For typesetting code
\usepackage{tcolorbox} % For creating colored boxes
\usepackage{courier} % For Courier New font
\usepackage{multirow}
\usepackage{graphbox}
\usepackage{booktabs}
\usepackage{amsmath}
\usepackage{amssymb}
\usepackage{graphicx}
\usepackage{subcaption}
\usepackage{tabularray}
\usepackage{caption}
\usepackage{url}
\usepackage{etoolbox}
\usepackage{wasysym}
\usepackage[export]{adjustbox}
\usepackage{tikz}
\usepackage{bm}
\usepackage{tocloft}
\usepackage[toc,page,header]{appendix}
\usepackage{minitoc}
\usepackage{titletoc}
\usepackage[accsupp]{axessibility}

\newtcolorbox{codebox}{
    colback=white, % Background color
    colframe=black, % Border color
    fontupper=\ttfamily, % Courier New font
    colframe=black!30, % Lighter border color (adjust the percentage as needed)
    arc=0pt, % Roundness of the corners
    boxsep=0pt, % Padding
    left=5pt, % Left margin
    right=5pt, % Right margin
}

\lstdefinestyle{mystyle}{
    basicstyle=\footnotesize\ttfamily, % Courier New font
    breaklines=true, % Allow line breaks
    breakatwhitespace=true, % Allow line breaks at whitespace
}

\definecolor{cvprblue}{rgb}{0.21,0.49,0.74}

\usepackage{cite}

\usepackage{academicons}
\usepackage{tikz}
\definecolor{orcidlogocol}{HTML}{A6CE39}
\definecolor{lime}{HTML}{A6CE39}
\DeclareRobustCommand{\orcidicon}{%
    \begin{tikzpicture}
    \draw[lime, fill=lime] (0,0) 
    circle [radius=0.16] 
    node[white] {{\fontfamily{qag}\selectfont \tiny ID}};
    \draw[white, fill=white] (-0.0625,0.095) 
    circle [radius=0.007];
    \end{tikzpicture}
    \hspace{-2mm}
}
\newcommand{\orcidWalter}{\href{https://orcid.org/0000-0003-4565-1272}{\orcidicon}}
\newcommand{\orcidRamandika}{\href{https://orcid.org/0009-0005-9321-9080}{\orcidicon}}
\newcommand{\orcidXingcheng}{\href{https://orcid.org/0000-0003-1178-5221}{\orcidicon}}
\newcommand{\orcidMingyu}{\href{https://orcid.org/0000-0002-8752-7950}{\orcidicon}}
\newcommand{\orcidKnoll
}{\href{https://orcid.org/0000-0003-4840-076X}{\orcidicon}}

\usepackage{multicol}
\usepackage{multirow}

\usepackage{algorithm}
\usepackage{algorithmic}

\usepackage[misc]{ifsym}

\makeatletter
\let\NAT@parse\undefined
\makeatother

\usepackage{hyperref} 
\hypersetup{
    bookmarks=false,
    colorlinks=true,
    breaklinks=true,
    linkcolor=blue,
    filecolor=magenta,      
    hypertexnames=true,
    urlcolor=cyan,
    pagebackref=true,
    pdftitle={WACV2025},
    pdfpagemode=FullScreen,
    citecolor=green
    }
\usepackage[capitalize]{cleveref}
\crefname{section}{Sec.}{Secs.}
\Crefname{section}{Section}{Sections}
\Crefname{table}{Table}{Tables}
\crefname{table}{Tab.}{Tabs.}

\def\wacvPaperID{1266} % *** Enter the WACV Paper ID here
\def\confName{WACV}
\def\confYear{2025}

\begin{document}

%%%%%%%%% TITLE - PLEASE UPDATE
\title{PointCompress3D: A Point Cloud Compression Framework for
Roadside LiDARs in Intelligent Transportation Systems}

% \author{First Author\\
% Institution1\\
% Institution1 address\\
% {\tt\small firstauthor@i1.org}
% % For a paper whose authors are all at the same institution,
% % omit the following lines up until the closing ``}''.
% % Additional authors and addresses can be added with ``\and'',
% % just like the second author.
% % To save space, use either the email address or home page, not both
% \and
% Second Author\\
% Institution2\\
% First line of institution2 address\\
% {\tt\small secondauthor@i2.org}
% }

\author{Walter Zimmer $^{\text{\Letter}~\bigstar}$\orcidWalter \qquad Ramandika Pranamulia$^\bigstar$\orcidRamandika\qquad Xingcheng Zhou~\orcidXingcheng\\ Mingyu Liu~\orcidMingyu \qquad Alois C. Knoll~\orcidKnoll\\\\
Technical University of Munich
% <-this % stops a space
%\thanks{The authors are with the School of Computation, Information and Technology, Technical University of Munich, 85748 Garching, Germany}
\setcounter{footnote}{-1}
\thanks{\text{\Letter}~Corresponding author: \tt\small walter.zimmer@tum.de}
\setcounter{footnote}{-1}
\thanks{$^\bigstar$These authors contributed equally.}
}

\makeatletter
\let\@oldmaketitle\@maketitle
\renewcommand{\@maketitle}{\@oldmaketitle
  \centering
  \vspace{-0.7cm}
  \url{https://pointcompress3d.github.io}\\[2pt]
  \includegraphics[width=1.0\linewidth,trim={0cm 0cm 0cm 0cm},clip]{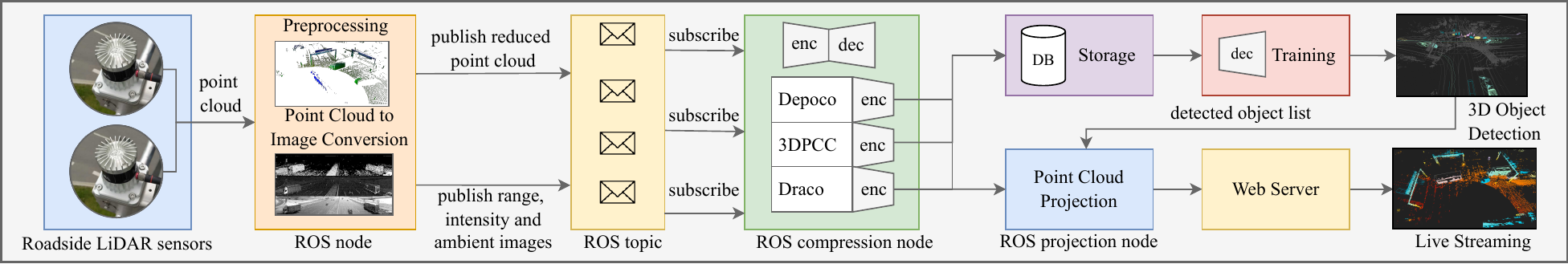}
  \captionof{figure}{\textbf{Visualization of the processing pipeline of our \textit{PointCompress3D} point cloud compression and streaming framework.} Our framework takes raw roadside LiDAR point clouds as input, processes them, and outputs compressed point clouds to facilitate downstream application tasks like 3D object detection, data storage, and real-time point cloud streaming on an ITS test bed for autonomous driving.}
  \label{fig:overview_figure}
  \vspace{0.4cm}
  }
\makeatother
\maketitle

%%%%%%%%% ABSTRACT
\begin{abstract}
   In the context of Intelligent Transportation Systems (ITS), efficient data compression is crucial for managing large-scale point cloud data acquired by roadside LiDAR sensors. The demand for efficient storage, streaming, and real-time object detection capabilities for point cloud data is substantial. This work introduces PointCompress3D, a novel point cloud compression framework explicitly tailored for roadside LiDARs. Our framework addresses the challenges of compressing high-resolution point clouds in real time while maintaining accuracy and compatibility with roadside LiDAR sensors. We adapt, extend, and integrate three cutting-edge compression methods and evaluate them on the real-world TUM Traffic datasets. Moreover, we deploy our compression framework on a real ITS test bed for autonomous driving and test it under real traffic conditions. After fine-tuning, we achieve a frame rate of 10 FPS while keeping compression sizes below 105 Kb, a reduction of 50 times, and maintaining object detection performance on par with the original data. In extensive experiments and ablation studies, we finally achieved a PSNR d2 of 94.46 and a BPP of 6.54 on the TUM Traffic datasets. The code is available on our project website.
\end{abstract}

%%%%%%%%% BODY TEXT

% TODO: add OctFormer (from 2023)

\begin{table*}[t] % Use table* instead of table for full-width
  \centering
  \resizebox{1.0\linewidth}{!}{
  \begin{tabular}{lllllllcl}
    \toprule
    \textbf{Method} & \textbf{Year} & \textbf{BPP}$\downarrow$ & \textbf{PSNR (d1) [db]$\uparrow$} & \textbf{PSNR (d2) [db]$\uparrow$} & \textbf{Enc. Speed [ms]$\downarrow$} & \textbf{Dec. Speed [ms]$\downarrow$} & \textbf{Code} & \textbf{Dataset}  \\
    \midrule
    MPEG-GPCC \cite{schwarz2018emerging}& 2018  & 1.00  \cite{cui2023octformer}  & 67.00 \cite{cui2023octformer} & 73.00 \cite{cui2023octformer} & 160.20 \cite{biswas2020muscle}& 112.80 \cite{biswas2020muscle} & \checkmark & SemanticKITTI %8iVSLF 
    \\
    \rowcolor{gray!10}
    Draco\cite{galligan2018draco} & 2018 & 5.98  \cite{he2022density} & -- & 44.19\cite{he2022density} & \,\,\,25.80 \cite{biswas2020muscle}& \,\,\,15.30 \cite{biswas2020muscle} & \textcolor{blue}{\checkmark} & - \\
    OctSqueeze \cite{huang2020octsqueeze} & 2020 & 1.00  \cite{cui2023octformer} & 69.00 \cite{cui2023octformer} & 74.00 \cite{cui2023octformer}  & \,\,\,35.75 & \,\,\,92.96 & \checkmark & SemanticKITTI%,NorthAmerica 
    \\
    MuSCLE \cite{biswas2020muscle}& 2020 & 5.00   & -- & \underline{81.00}  & \,\,\,20.80 & \,\,\,49.00 & - & SemanticKITTI \\
    % [] TODO: check encoding and decoding speed for VoxelContextNet. In [15] and [20] I cannnot find 0.43 ms encoding speed. In [20] they say 75 ms for decoding, not 419 ms.
    VoxelContext-Net \cite{que2021voxelcontext}& 2021 & 1.00  \cite{cui2023octformer} & \underline{72.00} \cite{cui2023octformer} & 76.00 \cite{cui2023octformer} & 388.90$^\dagger$ \cite{cui2023octformer} & 374.10$^\dagger$ \cite{cui2023octformer} & - & SemanticKITTI %,ScanNet  
    \\
    \rowcolor{gray!10}
    Depoco \cite{wiesmann2021deep}& 2021 & 4.98 \cite{he2022density} & -- & 40.01 \cite{he2022density} & \,\,\,32.00 \cite{he2022density}& \textbf{\,\,\,\,\,\,2.00} \cite{he2022density} & \checkmark & SemanticKITTI\\
    OctAttention \cite{fu2022octattention}& 2022 & 1.00  \cite{cui2023octformer} & \textbf{73.00} \cite{cui2023octformer} & 77.00 \cite{cui2023octformer} & \,\,\,\,\,\,\textbf{2.30}$^\ddagger$ \cite{cui2023octformer}& 2060.2$^\ddagger$ \cite{cui2023octformer} & \checkmark & SemanticKITTI \\
    GrASPNet \cite{pang2022grasp}& 2022 & 1.00  & 53.00 & 57.00 & 193.80  & 345.92 & \checkmark & Ford\\
    RIDDLE \cite{zhou2022riddle}& 2022 & 5.00  & -- & \textbf{95.00} & 532.51 & 966.30 & -- & SemanticKITTI\\
    D-PCC \cite{he2022density}& 2022 & 4.23 & -- & 47.98 & \,\,\,80.00 & \,\,\,24.00 & \checkmark & SemanticKITTI\\
    \rowcolor{gray!10}
    3DPCC \cite{beemelmanns20223d} & 2022 & 10.00 & -- & -- & \,\,\,35.00* & \,\,\,\,\,\,\underline{5.00}* & \checkmark & -- \\
    OctFormer \cite{cui2023octformer}& 2023 & 1.00 & \underline{72.00} & 76.00 & \,\,\,\,\,\,\underline{5.20} & \,\,\,\,\,\,7.30 & \checkmark & SemanticKITTI\\
    InterFrame \cite{akhtar2024inter}& 2024 & \textbf{0.06} & 70.10 & 73.00 & 364.00 & 714.00 & -- & 8iVFB v2\\
    \midrule
    \multicolumn{9}{l}{$^\dagger$Using a voxel size of 5. \hfill $^\ddagger$Using a context window size of N=1024.   \hfill*Using JPEG2000 compression instead of RNN.}\\
    %\multicolumn{9}{l}{}
  \end{tabular}
  } % end resizebox
  \caption{\textbf{Performance evaluation of state-of-the-art point cloud compression methods.} We report the \textit{BPP} metric for \textit{PSNR} d2 and highlight the supported methods in gray. Best results are marked in bold, second best are underlined.}
  \label{tab:sota_methods}
\end{table*}

\section{Introduction}
\label{sec:intro}
% \textcolor{red}{[important of PC compression \& framework necessity]}
%A project named \textit{AUTOtech.agil} is currently being undertaken by the German government in collaboration with research institutions and universities in Germany, as well as the industry, to achieve a reliable mobility system. 
This work is carried out within a project that contributes significantly to various aspects of improving mobility in transportation systems, including the creation of a live digital twin of the traffic to facilitate automated driving on level 4.
One concept for developing ITS systems to create live digital twins of the traffic involves the installation of cameras and LiDAR sensors at the sides of roads and junctions. 
These sensors capture data from the traffic, which is then fused, and disseminated to all entities who have subscribed to the live digital twin of the traffic. 
LiDAR sensors employ laser light for distance measurement, translate the environment into a collection of 3D points, and facilitate the creation of precise three-dimensional digital twins of the surroundings. 
These high-resolution sensors operate at 10-30 Hz and emit up to 2.6 Mio. points per second (pps). This results in a significant amount of data, ranging from 5-25 MB per frame.
% \textcolor{red}{[smooth transition = challenge for transferring this massive point cloud + importance of compression technologies + importance of our work]}
There are many use cases for roadside infrastructure LiDAR sensors. One of the most crucial tasks is to process point cloud data to detect traffic participants. Numerous scene understanding \cite{zhou2023vision} and object detection methods \cite{zimmer2023real,zimmer2023infradet3d,zimmer2022realdomain,zimmer2022survey} have been developed for that purpose, using, e.g., \textit{PointPillars} \cite{lang2019pointpillars} as their baseline. In other cases, they can be incorporated with multiple input sources like images to perform multi-modal object detection \cite{li2022deepfusion,melotti2020multimodal,ghita2024activeanno3d,hekimoglu2023multi}. However, the large amount of data poses challenges to processing, storing, and transmitting these point clouds.

\paragraph{Our contribution is the following:}
\begin{itemize}
    \item We propose a point cloud compression framework shown in \cref{fig:overview_figure} for roadside infrastructure LiDARs.% sensors and a dev-kit.
    \item We provide an in-depth comparison and analysis of state-of-the-art compression methods on  \textit{SemanticKITTI} \cite{behley2019semantickitti}, \textit{Ford} \cite{agarwal2020ford} and the \textit{TUM Traffic} datasets \cite{cress2022a9,zimmer2023tumtraf,cress2024tumtraf,zimmer2024tumtrafv2x}.
    \item We extend existing compression methods to make them compatible with our roadside LiDAR sensors.
    % [] TODO: what method (depoco, draco, 3DPCC) was used to get these results?
    % [] TODO: on what dataset do we get these results (TUMTraf-I or TUMTraf-V2X)?
    \item We perform extensive experiments and ablation studies on the \textit{TUMTraf Intersection} and \textit{TUMTraf V2X Cooperative Perception} dataset and achieve a \textit{PSNR d2} of 94.46 and a \textit{BPP} of 6.54. 
    % [] TODO: for final submission: add link to dev-kit
    %using our dev-kit\footnote{\url{https://github.com/tum-traffic-dataset/tum-traffic-dataset-dev-kit}}.
    \item We open-source the code of our framework, which contains the point cloud projection and compression module, and provide a website with video results.
\end{itemize}

\section{Related Work}
\label{sec:related-works}
% [x] TODO: \cite{beemelmanns20223d}
% [] TODO: \cite{galligan2018draco} -> draco
Point cloud data consists of $(x,y,z)$ geometric coordinates and can also contain additional attributes based on the manufacturer, such as intensity, reflectivity, time, ambient, ring, and range. Moreover, attributes such as RGB colors or normals can be added to the point clouds. However, most point cloud compression methods focus only on geometry compression. Methods compressing attributes in addition to geometry do both compressions separately. Geometry compression is highly prioritized because it is still difficult to get a good balance of fast compression speed, low bit rate, and low reconstruction error. 

\begin{figure*}[t]
    \centering
    %\captionsetup{type=figure}
    %\includegraphics[width=.325\textwidth,trim={0cm 1cm 0cm 0cm},clip,frame]{figs/side_ori.png}
    %\includegraphics[width=.325\textwidth,trim={0cm 1cm 0cm 0cm},clip]{figs/side_e0.png}
    %\includegraphics[width=.325\textwidth,trim={0cm 1cm 0cm 0cm},clip,frame]{figs/side_e1.png}
    %\includegraphics[width=.3\textwidth]{figs/side_e2.png}
    %\includegraphics[width=.325\textwidth,trim={0cm 1cm 0cm 0cm},clip,frame]{figs/side_e3.png}
    \includegraphics[width=1.0\textwidth,trim={0cm 0cm 0.1cm 0cm},clip,frame]{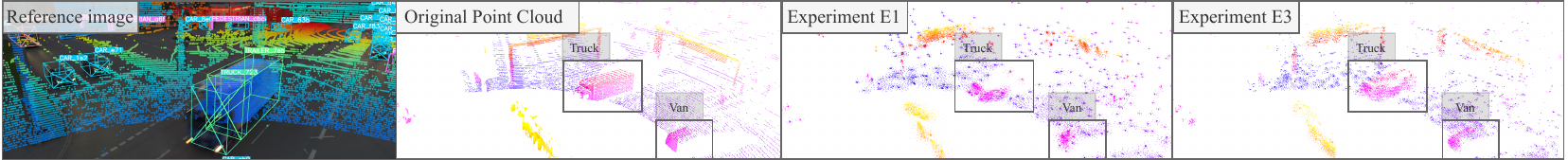}
    \caption{\textbf{Visualization of ablation study results for \textit{Depoco} on the \textit{TUMTraf V2X Cooperative Perception} \cite{zimmer2024tumtrafv2x} dataset.} From left to right: a) Reference image b) View of the original point cloud. b) Experiment E1: Reconstructed point cloud with a subsampling distance of 3.0 and a minimum kernel radius of 1.5. c) Experiment E3: Reconstructed point cloud with a subsampling distance of 1.0 and a minimum kernel radius of 1.2. Both reconstructed point clouds are generated with max. 30,000 points and a grid size of 40x40x15 m.}
    \label{fig:confiruation-quality}
\end{figure*}

\subsection{Geometry compression} 
The focus is on compressing 3D coordinates, representing geometrical shapes, with various techniques. The first step usually involves converting 3D coordinates into a dense representation, such as tree structures \cite{schwarz2018emerging, galligan2018draco}, voxels \cite{que2021voxelcontext, pang2024pivot}, or 2D images \cite{beemelmanns20223d, houshiar20153d}. After obtaining this dense representation, various methods can be applied depending on the previously chosen structure. In general, the processing methods will exploit the statistical properties of the data using hand-crafted or learning-based methods.

\subsection{Octree representation} 
We can encode the \textit{octree's} occupancy nodes using predictive and arithmetic coding for the \textit{octree} representation. Predictive coding is used to predict the occupancy symbol of the \textit{octree} nodes based on their neighbors. This can be achieved by hand-crafted entropy models \cite{schwarz2018emerging} or by using deep learning-based methods to automatically learn the entropy model \cite{tatarchenko2017octree, huang2020octsqueeze, fu2022octattention}. Arithmetic coding is mostly used to encode the residual error from comparing the original data to the reconstructed data \cite{quach2022survey}. Unfortunately, this approach is highly complex when dealing with large point cloud data since constructing the \textit{octree} and performing traversal operations can be time-consuming. A precision loss might also be part of the \textit{octree}-based codec because of the quantization process that took place. We might minimize \textit{octree} complexity and the precision loss by limiting the \textit{octree} depth and using a less aggressive quantization level, respectively. Still, it results in a lower compression ratio. 

% \begin{figure}[t]
%   \centering
%  \includegraphics[width=0.7\columnwidth]{figs/quality_graph.png}
% \caption{Comparing several compression methods for KITTI dataset using BPP as efficiency metric and Point to Point PSNR as effectivity metric \cite{huang2020octsqueeze}}
% \label{fig:quality-graph}
% \end{figure}
\begin{algorithm}
\caption{Voxelization Process}
\label{alg:voxelization}
\begin{algorithmic}[1]
    \STATE \textbf{Input:} Point cloud data $\mathcal{P}$, Voxel size $v$
    \STATE \textbf{Output:} Voxel grid representation $\mathcal{V}$
    \STATE 
    \STATE \textbf{Procedure:}
    \STATE Define the resolution of the voxel grid by specifying the voxel size $v$.
    \STATE Divide the 3D space occupied by the point cloud $\mathcal{P}$ into a regular grid of voxels.
    \FOR{each voxel $voxel$ in $\mathcal{V}$}
        \STATE Initialize voxel values based on assignment method:
        \STATE \quad \textbf{Binary occupancy:} Set voxel value to 1 if it contains one or more points, 0 otherwise.
        \STATE \quad \textbf{Averaged properties:} Calculate average properties (e.g., intensity, range) of points within voxel.
        \STATE \quad \textbf{Density:} Count the number of points within the voxel and store them as a value.
    \ENDFOR
\end{algorithmic}
\end{algorithm}

\subsection{Voxelization} 
Voxelizing the point cloud into 3D structural grids with specified grid sizes as shown in \cref{alg:voxelization} is another common representation. This extends the concept of 2D images constructed by pixels, while in this case, the 3D space is made of \textit{voxels}. This aims to bring the concepts of 2D processing into 3D space, one of which is a \textit{3D CNN autoencoder} that we can employ to compress those \textit{voxels} through the encoder part and decompress them through the decoder part \cite{que2021voxelcontext}. Another method we can apply to those \textit{voxel} representations is masked convolutions to predict \textit{voxel} occupancy probabilities sequentially, feeding previously predicted \textit{voxels} as context \cite{nguyen2021learning}, which is inspired by combining \textit{voxels} and the \textit{octree} representation. This voxelization method has a similar problem to that of \textit{octree} methods, where the size of the \textit{voxels} will affect the compression speed.

\subsection{3D Transformation} 
As for projection transformation, where the transformation results are images, compression technology for images has advanced significantly. Hence, we can utilize available technologies such as \textit{JPEG} or deep learning-based image compression \cite{toderici2017full} methods to get a good compression result like in \cite{beemelmanns20223d}. The advantage of this method is that the transformation directly reduces the size without the compression process. Each 3D point represented as $(x,y,z)$ gets reduced to a single variable representing the range of the point from the observation source. Beemelmanns et al. \cite{beemelmanns20223d} achieve a lossless compression and precisely reconstruct the original point.
Nevertheless, this approach is limited to a specific scenario in that no 3D point is occluded by another point. Merged point clouds from multiple LiDARs cannot be reconstructed this way.
It also relies on the projection parameters, such as the position of the sensor and the azimuth angles, which are specific to the LiDAR type and manufacturer. Moving the sensor will require a recalibration of the LiDAR.

\section{Methodology}
We first compare 13 available state-of-the-art methods and analyze their performance based on the BPP and PSNR metrics, as well as the encoding and decoding speed.
Then, we choose the best three state-of-the-art compression algorithms (\textit{Depoco}, \textit{3DPCC}, and \textit{Draco}) and adapt them to make them compatible with the \textit{TUMTraf} datasets \cite{liu2024survey} and the Ouster \textit{LiDAR} sensor. 
Furthermore, we extend these algorithms and fine-tune the parameters to find the best tradeoff between compression ratio and speed and improve the compression quality.
Finally, we integrate them into our \textit{PointCompress3D} compression and streaming framework and connect them to a LiDAR 3D object detector to stream compressed point clouds with detected 3D objects.
% and integrate it deploy it on our live system.

\subsection{Method Selection}
To choose the most promising method from the state of the art, we first evaluate and inspect various aspects of each method, such as compression efficiency, compression effectiveness, compression speed, and code availability (see Table \ref{tab:sota_methods}). 
% LIMITATION
%Unfortunately, building Table \ref{tab:example} is not straightforward since there are issues when gathering and standardizing information about bitrate, PSNR, and speed from various different papers. 
% LIMITATION
%but it is impossible for us to plot all the methods, and no paper has ever done that before. 
The bitrate indicates the number of bits required per data point in a point cloud. 
Factors like spatial information and additional data stored influence it. 
Each bitrate corresponds to a unique \textit{Peak Signal-to-Noise Ratio (PSNR)} value. 
% LIMITATION
Still, variations in the \textit{PSNR} formula, particularly the \textit{MAX} parameter, can affect comparisons between methods. 
% LIMITATION
%Due to hardware specifications and implementation details, speed is not directly comparable across methods. 
% LIMITATION
%Additionally, some metrics mentioned may not be consistent or available in other sources.

%Despite all the inconsistencies mentioned above, we still need to be able to identify promising methods before proceeding with our internal dataset evaluation. Therefore, 

The strategies employed to ensure comparability between methods include:
\begin{itemize}
     % LIMITATION -> only geometry
    \item We only focus on geometry compression, i.e., spatial attributes.
    % LIMITATION -> only SemanticKITTI
    \item As \textit{SemanticKITTI} is the most widely used dataset for point cloud compression, we focus on gathering methods evaluated on this dataset.
    % LIMITATION -> only SemanticKITTI
    \item We only consider methods that use 1 or 5 \textit{BPP} for each \textit{PSNR} value evaluation.
    % LIMITATION -> it is not complete
    \item When the performance of a given method is not given, we refer to the related work that states the desired metric to provide cross-sourced metric findings.
    % LIMITATION -> only bits per point considered
    \item We only consider \textit{bits per point} as required to encode spatial attributes.
\end{itemize}

% LIMITATION -> potentially incorrect conclusions
%Although all the strategies above do not rule out the possibility of drawing incorrect conclusions about which methods are the best, at least the strategies mentioned can reduce the likelihood of drawing incorrect conclusions.

Looking at Table \ref{tab:sota_methods}, with priority given to speed and code availability before reconstruction error, we find four promising methods that run under 10 FPS. Although \textit{MuSCLE} \cite{biswas2020muscle} fulfills the speed requirement, the code is not publicly available. Of all four candidates, \textit{Depoco} is the fastest one, followed by \textit{Draco}, with their total encoding and decoding speed being 34 ms and 40 ms, respectively. \textit{Depoco's} and \textit{Draco's} reconstruction errors are also comparable and not significantly different. Hence, we evaluate both \textit{Draco} and \textit{Depoco} alongside another additional method, \textit{3DPCC-RNN} \cite{beemelmanns20223d}, on our dataset. \textit{3DPCC-RNN} converts 3D point clouds to range images and vice versa. This enables us to utilize image compression algorithms, which are more advanced and promising than directly compressing the 3D point cloud.

\subsection{Extensions}
We extend and modify three point cloud compression algorithms to make them compatible with roadside LiDAR sensors and improve their compression quality and speed.

\begin{figure}[t]
    \centering
    %\captionsetup{type=figure}
    % \includegraphics[width=.49\linewidth] {figs/side_e3.png}
    %\includegraphics[width=.49\linewidth,trim={0cm 1cm 0cm 0cm},clip]{figs/min_kernel_0.5x.png}\hskip 0.2ex
    %\includegraphics[width=.49\linewidth,trim={0cm 1cm 0cm 0cm},clip]{figs/min_kernel_2x.png}
    \includegraphics[width=1.0\linewidth]{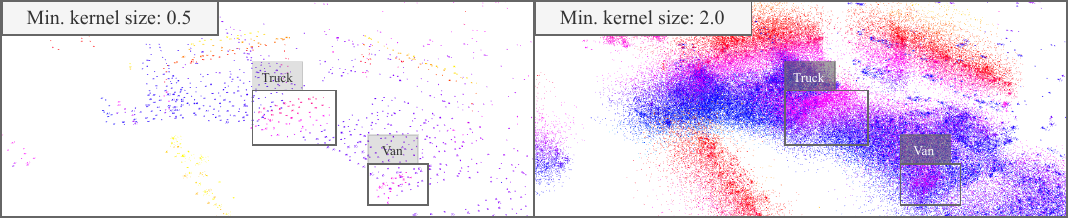}
    \caption{\textbf{Ablation study results for \textit{Depoco}:} Adjusting the \textit{minimum kernel radius} parameters by 0.5 times and 2 times the original value, respectively.}
    \label{fig:min_kernel_radius}
\end{figure}

\textbf{Depoco.} \textit{Depoco} is a learning-based compression method using autoencoders \cite{wiesmann2021deep}. Configuration files and pre-trained weights of the \textit{Depoco} model are provided. 
%Some important files within its code repository are \texttt{encode.py}, \texttt{decode.py}, \texttt{trainer.py}, and \texttt{submap\_handler.py}. 
%They store all configuration files under \texttt{network\_files/} folder, there are 4 provided configuration files namely \texttt{e0.yaml}, ..., \texttt{e3.yaml}. 
Using grid search, we tune the parameters listed in \cref{tab:depoco-tune} and start from the configuration file with the best qualitative result. Figure \ref{fig:confiruation-quality} shows how the initial configuration performs on the \textit{TUMTraf Intersection} dataset. 
We can see that the configuration with a subsampling distance of 1.0 and a minimum kernel radius of 1.2 performs best. 
Both compressed images are generated with max. 30,000 points and a grid size of [40, 40, 15] m. 
The truck and the car look almost identical as in the original point cloud frame. 
The mean encoding time for experiment E1 is 64 - 65 ms, and for experiment E3 is 110 - 120 ms, while the mean decoding time for both configurations is 90 ms. 
% [] TODO: why not starting from E1 if the encoding time for E1 is smaller (64 ms)?
Therefore, we opted to start tuning from configuration E3, prioritizing encoding speed within the 0.1 s limit while ensuring superior quality performance.\\
% [] TODO: get speed for E1 and mean decoding time for E1+E3.
%e2 is 0.13 - 0.14 s and e3 is 0.11 - 0.12 s while the decoding meantime for both configurations is 0.003 s. 
\begin{figure}[t]
    \centering
    \includegraphics[width=1.0\linewidth]{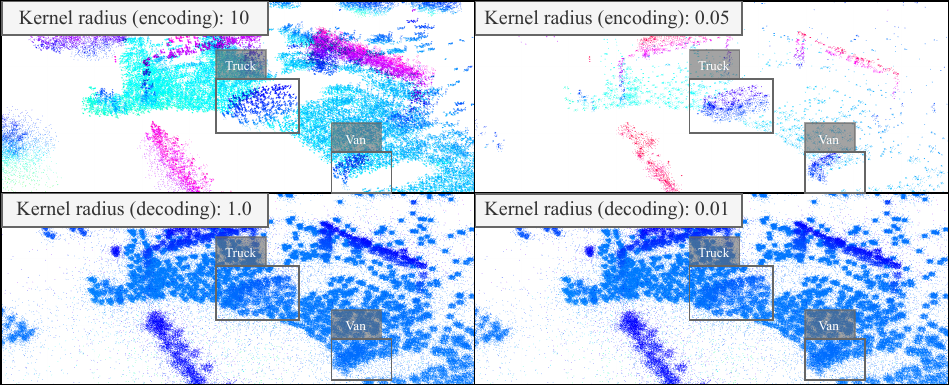}
    \caption{\textbf{Ablation study results for \textit{Depoco}:} We set the \textit{kernel radius} in encoding block to 10 and 0.05 (top row) while keeping the decoding block values constant at 0.05. Subsequently, we set \textit{kernel radius} in the decoding block to 1 and 0.01 (bottom row) while keeping the encoding block values constant at 1.0.}
    \label{fig:kernel_radius}
\end{figure}
\begin{figure*}
    \centering
    \includegraphics[width=1.0\textwidth,frame]{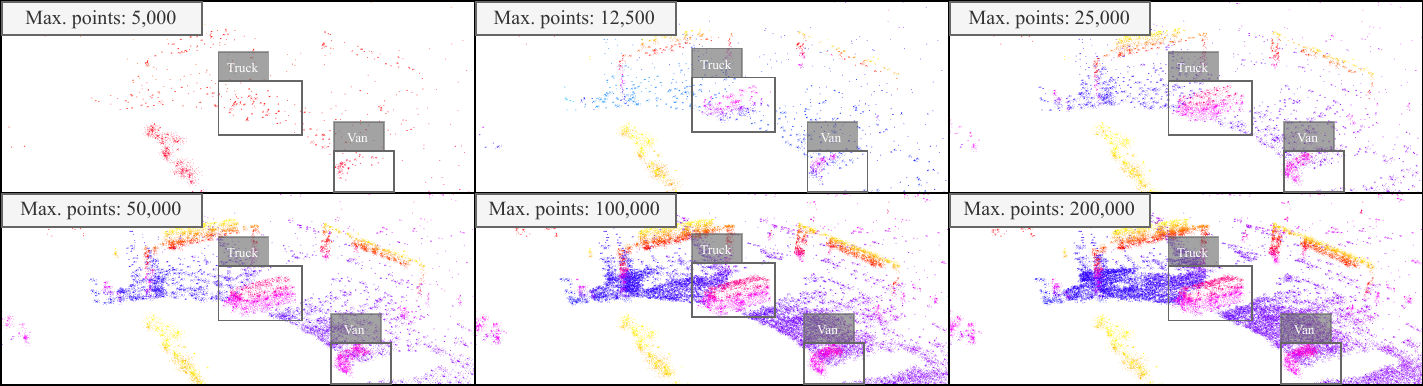}
    \caption{\textbf{Visualization of the ablation study results for \textit{Depoco} for the max. number of points}. The point cloud shows a side view of a truck and a car with 5,000, 12,500, 25,000, 50,000, 100,000, and 200,000 points. }
    \label{fig:max_nr_pts}
\end{figure*}
We begin by tuning the following parameters: \textit{subsampling distance}, \textit{min. kernel radius}, and \textit{kernel radius}. 
To maintain compatibility with the provided pre-trained weights, we retain the original \textit{subsampling distance}. 
Adjustments to this parameter were found to cause block dimension mismatches. Thus, it remains unchanged to utilize the existing weights effectively.\\ 
Next, we adjust the \textit{min. kernel radius} parameter. 
The result is depicted in Figure \ref{fig:min_kernel_radius}. 
We observe that enlarging the variable results in denser points while reducing it decreases the number of points. 
% LIMITATION -> current value is the optimal one
However, our current value appears optimal, as the object boundaries are well-defined and perceivable. 
Moving on, we tune the \textit{kernel radius} in encoder and decoder blocks, with the default settings being 1.0 for the encoder and 0.05 for the decoder. 
% LIMITATION -> original parameters yield the best results
Once again, examining the results in Figure \ref{fig:kernel_radius}, we find that the original parameter yields the best result. 
Subsequently, we adjust the \textit{max. number of points}, similar to the \textit{min. kernel radius} and \textit{kernel radius}. 
Notably, for \textit{max. number of points}, the points added to the scene are around the dynamic objects, where the object remains perceptible but becomes densely populated with points, lacking inner details shown in Figure \ref{fig:max_nr_pts}. 
We do not inspect the \textit{grid feature dimension} because the grid size is the tile unit where compression is applied. 
This implies that smaller grid sizes should result in better compression quality. 
Moreover, the \textit{grid feature dimension} does not affect the reconstruction quality. It is only modified if additional attributes, such as the intensity, need to be encoded.

\textbf{3DPCC.} In \cite{beemelmanns20223d}, Beemelmanns \textit{et al.} present a novel approach for a lossless and calibrated 3D-to-2D transformation which allows compression algorithms to efficiently exploit spatial correlations within the 2D representation. The authors employ \textit{JPEG2000}, \textit{PNG}, and \textit{RNN}-based compression methods \cite{toderici2017full} to compress the images. We use \textit{Ouster} LiDAR sensors, which require specific modifications due to differences in settings compared to \textit{Velodyne} LiDARs.

It's important to note that \textit{Velodyne's} point cloud data does not come with range images, necessitating a conversion algorithm. In contrast, \textit{Ouster} LiDARs provide range images alongside point cloud data. However, without azimuth maps, we are restricted to using the built-in range images directly. In some experiments, we replaced the point cloud-transformed range image with the built-in range image for the reconstruction, as seen in Tables \ref{table:image-compression-size} and \ref{table:image-compression-error}.
%for comparison in the evaluation section.

% Now, looking at the code repository, some important files are 
% \begin{itemize}
%     \item \texttt{pointcloud\_to\_rangeimage\_node.cpp}
%     \item \texttt{compression.launch}
%     \item \texttt{rangeimage\_to\_pointcloud\_node.cpp}
%     \item \texttt{pointcloud\_to\_rangeimage.yaml}
% \end{itemize}. 

The \textit{3DPCC} compression framework is integrated into a \textit{ROS} system. Communication occurs through \textit{ROS} topics and messages. 
To utilize point clouds from Ouster LiDARs, we had to convert them into the correct point cloud data format.
%replacing the message from the ROS topic. 
%This modification is straightforward. 
%Later, instead of sending the images to another \textit{ROS} topic, we store them in the filesystem. 
%These modifications are implemented within `pointcloud\_to\_rangeimage\_node.cpp`. 
Three images representing one point cloud frame are produced: the range, intensity, and azimuth images. To convert them back to the point cloud structure, we perform a 2D to 3D transformation.
%and storing them back to the file system as \textit{PCD} point cloud files. 
%This process occurs in `rangeimage\_to\_pointcloud\_node.cpp`.

Next, we adjust the azimuth and elevation angles in 
%\texttt{pointcloud\_to\_rangeimage.yaml} 
the configuration file because the \textit{Ouster} LiDAR mounted on our roadside infrastructure has a below-horizon beam configuration.
\textit{Velodyne} LiDAR sensors used in \cite{beemelmanns20223d} have 32 beams, 16 above the horizon and 16 below. Our \textit{Ouster} LiDAR sensors have 64 beams, all set up to point below the horizon. We also need to order the elevation angle from lower to higher, and the azimuth angle should correspond to the right elevation angle representing a configuration of a laser beam.

In the last step, we modify the input point cloud. 
The initial order of points is sorted based on width, which is 2048 in our case. 
Therefore, the first 2048 rows correspond to one LiDAR beam closest to the horizon. 
The next 2048 rows correspond to the second LiDAR beam closest to the horizon. 
This is equivalent to all 64 beams. 
Hence, we have 2048 * 64 = 131,072 points in a single point cloud scan. 
We change the order of the points to columns: The first column represents the different 64 LiDAR scans, and the second column represents another 64 LiDAR scans, up to 2048 columns scanned.

\textbf{Draco.} To use the \textit{Draco} point cloud compression method in our framework, we integrate the \textit{DracoPy} \textit{Python} bindings \cite{dracopy}, which wrap the Draco executables. We optimize the quantization bits and compression level parameters specifically for \textit{Ouster} LiDAR data. 
% [] TODO: what are the final fine-tuned parameter values? How many quantization bits work best? what compression level works best?
By fine-tuning these parameters using grid search, we ensured an efficient and effective compression of point clouds generated from \textit{Ouster} LiDARs to decrease the file storage size and improve transmission bandwidth for LiDAR-based applications.

%\twocolumn[{%
%\renewcommand\twocolumn[1][]{#1}%
%\maketitle
% \begin{figure*}
%     \centering
%     %\captionsetup{type=figure}
%     \includegraphics[width=.5\textwidth]{figs/lidar_different_position.png}
%     % \includegraphics[width=.3\textwidth]{figs/lidar_on_road.png}
%     %\captionof{figure}{Positional and angle differences on LiDARS setup in Velodyne and our Ouster}
%     \label{fig:lidar-different-position}
% \end{figure*}%
%}]
\section{Evaluation}
To gain more insights into the performance of the selected compression methods, we perform several experiments and ablation studies on the \textit{TUMTraf Intersection} and the \textit{TUMTraf V2X Cooperative Perception} dataset. 

\begin{table*}[t] % Use table* instead of table for full-width
  \centering
  \resizebox{\linewidth}{!}{
  \begin{tabular}{lcrrrrrrrr}
    \toprule
    \textbf{Grid Size} & \textbf{Max. Points} & \textbf{Enc. (ms)}$\downarrow$ & \textbf{Dec. (ms)}$\downarrow$ & \textbf{Enc. VRAM (GB)}$\downarrow$ & \textbf{Dec. VRAM (GB)}$\downarrow$ & \textbf{PSNR d1}$\uparrow$ & \textbf{PSNR (d2)}$\uparrow$ & \textbf{BPP}$\downarrow$ & $\mathbf{mAP_{3D}}$$\uparrow$  \\
    \midrule
    \multirow{3}{*}{8x8x3}
    & 50,000  & \textbf{220} & 2.7 & \textbf{3.9} & \textbf{3.9} &  15.32 & 23.68 & \textbf{7.48} & 13.32\\
    & 100,000  & 410 & 2.7  & 6.0 & 4.3 & 21.81 & 30.13 & 11.72 & 17.29 \\
    & 200,000  & 520 & 2.8 &  5.5 & 7.1  & \textbf{24.18} & \textbf{32.88} & 13.57 & \textbf{19.39}\\
    \midrule
    \multirow{3}{*}{16x16x6}
    & 50,000  & 230 & 2.6 & \textbf{3.0} & \textbf{2.7} & -7.81 & -0.19 & \textbf{5.17} & 12.21  \\
    & 100,000  & 350 & \textbf{2.5} & 4.3 & 3.7 & -3.59 & 3.87 & 7.03 & 19.50 \\
    & 200,000  & 510 & 2.7 & 6.8 & 4.0 & \textbf{-1.46} & \textbf{6.31} & 7.73 & \textbf{20.91} \\
    \midrule
    \multirow{3}{*}{24x24x9} & 50,000  & 240 & \textbf{2.5}  & \textbf{2.2} & \textbf{2.7} & -5.57 & 0.99 &  \textbf{3.90} & 13.59 \\
    & 100,000  & 410 & 2.6 & 5.5 & 3.1 & -1.87 & 5.50 & 4.94 & \textbf{19.75}\\ 
    & 200,000  & 530 & 2.7 & 6.6 & 3.2 & \textbf{-0.18} & \textbf{7.47} & 5.31 & 19.32 \\
    \bottomrule
  \end{tabular}
  }% end resizebox
    \caption{\textbf{Parameter tuning results for \textit{Depoco} on the max. number of points and the grid size.} We use the \textit{TUMTraf Intersection} dataset to find the best parameters and evaluate the compression method on the \textit{TUMTraf V2X Cooperative Perception} dataset.}
  \label{tab:depoco-tune}
\end{table*}

\subsection{Performance Metrics}
In compression, we encounter diverse metrics assessing performance and quality. We balance speed, quality, and performance and determine the best trade-off.
%yet lack a unified metric for comparison, unlike object detection. 
We assess the reconstruction quality at specific compression rates, comparing metric graphs to determine method superiority.

\begin{table}[b]
    \centering
    \setlength{\tabcolsep}{3pt}
    \resizebox{\columnwidth}{!}{
    \begin{NiceTabular}{lrrrrr}
    \CodeBefore
    \rectanglecolor{gray!10}{5-1}{5-6}
    \Body
    \toprule
     \textbf{Input Image} & \textbf{Size (gen.)} & \textbf{Size (orig.)} & \textbf{TinyJPG}& \textbf{PNG} & \textbf{ImgMagic}\\
    \midrule
    Range image & 113.70 & 43.50 & 13.40 & 111.70 & 92.10 \\
    Azimuth image & 56.30 & -  & 10.20 & 65.30 & 56.30 \\
    Intensity image & \textbf{20.72} & - & \textbf{6.20} & \textbf{21.90} & \textbf{20.72} \\
    \midrule
    \textbf{Total Size (KB)} & 190.70 & 120.50 & 29.80 & 204.90 & 169.10
    % \\
    % \hline
    % \\
    % PSNR d1 (dB) & 37.247 / 41.45 & 41.45 & -
    \\
    \bottomrule
    \end{NiceTabular}
    }
    \caption{\textbf{Results of the point cloud image (range, azimuth, and intensity) compression of \textit{3DPCC} on the \textit{TUMTraf-I} dataset.} Size is given in KB.}
    \label{table:image-compression-size}
\end{table}

\textbf{PSNR.} Following \cite{javaheri2020improving} based on \cite{schwarz2018emerging}, we use \textit{PSNR} (Peak Signal-to-Noise Ratio) with the \textit{d1} and \textit{d2} metric, as reconstruction metric. \textit{PSNR} is a widely used metric in image and video compression to evaluate the quality of the compressed image or video compared to the original ones. 
It measures the reconstruction quality in both directions by comparing the \textit{mean squared error} (MSE) between the original and compressed point cloud (see \cref{eq:psnr}).
The PSNRs of the two directions are then combined to obtain a single symmetric PSNR value with the maximum pooling function, as defined in \cref{eq:psnr_combined}.
\vspace{-0.45cm}
\begin{equation}\label{eq:psnr}
    \text{PSNR}_{A,B} = 10 \cdot \log_{10} \left( \frac{{\text{p}^2_s}}{{d^{MSE}_{A,B}}} \right)
\end{equation}
\begin{equation}\label{eq:psnr_combined}
    \text{PSNR} = max(\text{PSNR}_{A,B}, \text{PSNR}_{B,A})
\end{equation}
where $p$ is the signal peak $d$ is the average squared error (i.e. MSE) between all point in point cloud A and their corresponding neighbor in point cloud B.
\textit{PSNR} is measured in decibels (dB), and a higher \textit{PSNR} value indicates better quality because the \textit{MSE} is smaller. In the case of point clouds, the \textit{MAX} value can be set in various ways, depending on the compression type, such as voxelized or not \cite{javaheri2020improving}. 
The most common method is to set it to the value of the maximum diagonal of the box bounding of the point cloud.

\begin{table}[b]
    \centering
    \resizebox{\columnwidth}{!}{
    \begin{tabular}{lrrr}
    \toprule
    \textbf{Input Representation} & \textbf{PSNR d1} & \textbf{PSNR d2} & \textbf{mAP@50} \\
    \midrule
    Generated RIA$^{\mathrm{\dag}}$ & 25.93 & \textbf{36.23} & \textbf{7.49} \\
    Generated IA + Ori. R$^{\mathrm{\dag}}$ & 21.57 & 24.87 & 0.00 \\
    Generated IA + Comp. R$^{\mathrm{\dag}}$(ImageMagick) & \textbf{26.38} & 33.72 & 2.48 \\
    Comp. RIA$^{\mathrm{\dag}}$ (TinyJPEG) & 26.10 & 26.92 & 0.04 \\\bottomrule
    \multicolumn{3}{l}{\scriptsize{$^{\mathrm{\dag}}$A=azimuth, I=intensity, R= range image.}} \\
    \end{tabular}
    }% end resizebox
    \caption{\textbf{Evaluation of the \textit{3DPCC} compression and the \textit{PointPillars} 3D object detection performance on the \textit{TUMTraf-I} dataset.} We compare four different input representations.}
    \label{table:image-compression-error}
\end{table}

\begin{figure*}[t]
    \centering
    \includegraphics[align=c,width=1.0\textwidth]{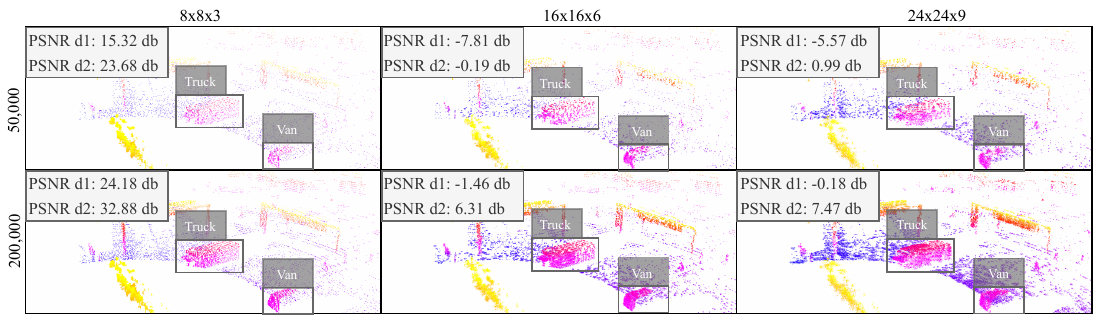}
    \caption{\textbf{Visualization of the ablation study results for \textit{Depoco}.} We show the reconstructed point clouds for different grid sizes (x-axis) and point sizes (y-axis) on the \textit{TUMTraf V2X Cooperative Perception} \cite{zimmer2024tumtrafv2x} dataset and report the PSNR d1 and d2 metrics.}
    %\caption{Visualization of the ablation study results. Reconstructed point cloud result for different values of the max. number of points and grid size.}
    \label{fig:depoco-params}
\end{figure*}%

\begin{figure*}[bht!]
    \centering
    \includegraphics[width=1.0\linewidth,frame,clip,trim={0cm 0cm 0.1cm 0cm}]{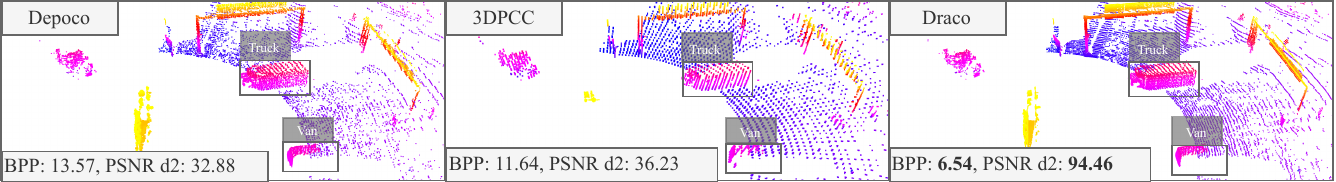}
    \caption{\textbf{Qualitative reconstruction results.} Qualitative comparison of \textit{Depoco}, \textit{3DPCC} and \textit{Draco} on the \textit{TUMTraf V2X Cooperative Perception} \cite{zimmer2024tumtrafv2x} dataset. \textit{Draco} outperforms \textit{Depoco} and \textit{3DPCC} regarding the \textit{BPP} and \textit{PSNR d2} metrics. \label{fig:depoco_3dpcc_draco_on_tumtraf}}
\end{figure*}

\textbf{Chamfer Distance.} It's similar to \textit{PSNR} in terms of a measurement system for reconstruction error. The \textit{Chamfer} distance between two point clouds $P_1=\left\{\bm{x_i} \in \mathbb{R}^3\right\}_{i=1}^n$ and $P_2=\left\{\bm{x_j} \in \mathbb{R}^3\right\}_{j=1}^m$ is defined as the average distance between pairs of nearest neighbors between $P_1$ and $P_2$ i.e.
\begin{equation}
\begin{split}
\operatorname{C}\left(P_1, P_2\right)=\frac{1}{2n} \sum_{i=1}^n \left| \bm{x_i} - \mathrm{NN}\left(\bm{x_i}, P_2\right) \right| + \\
\frac{1}{2m} \sum_{j=1}^m \left| \bm{x_j} - \mathrm{NN}\left(\bm{x_j}, P_1\right) \right|
\end{split}
\end{equation}
where $\mathrm{NN}(\bm{x}, P)=\operatorname{argmin}_{\bm{x^{\prime}} \in P}\left\|\bm{x}-\bm{x^{\prime}}\right\|$ is the nearest neighbor function and $n, m$ are the number of points in $P_1$, $P_2$ respectively. The \textit{Chamfer} distance has a minimum of 0, meaning that two sets are spatially identical. We can also adjust the distance metric $||\cdot||$ with various distance metrics such as \textit{Manhattan} or \textit{Euclidian} distance metric.

\begin{figure*}[t]
    \centering
    \includegraphics[width=0.32\linewidth,trim={0cm 0cm 0cm 0.7cm},clip]{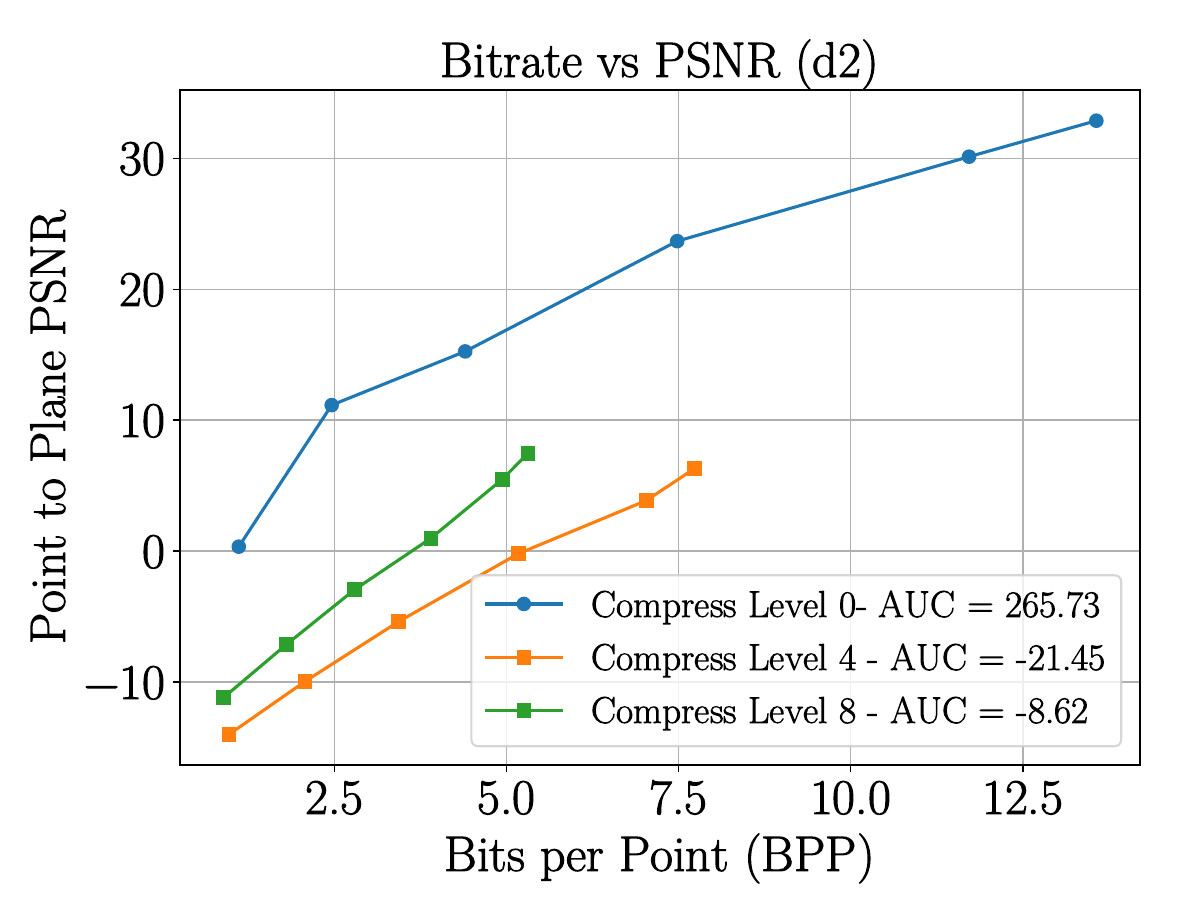}
    \includegraphics[width=0.32\linewidth,trim={0cm 0cm 0cm 0.7cm},clip]{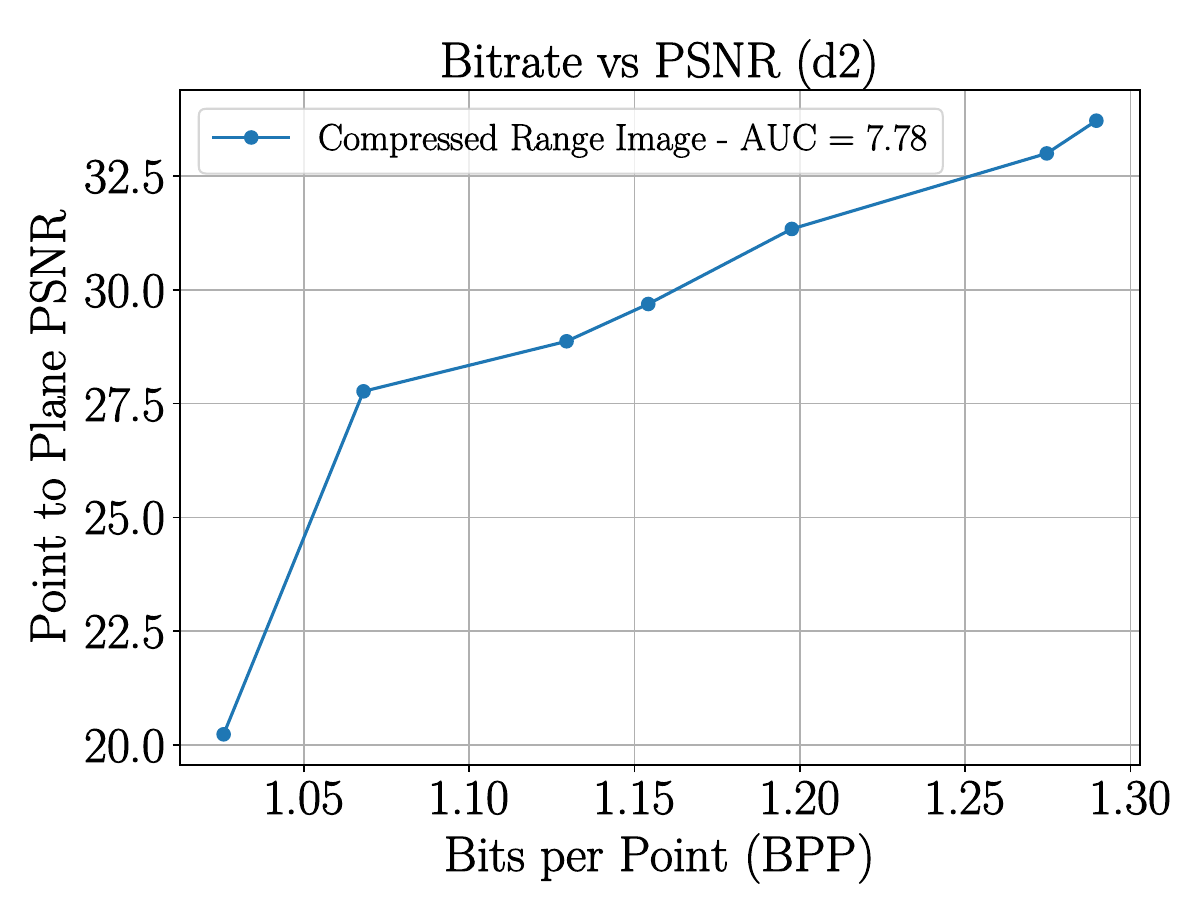}
    \includegraphics[width=0.32\linewidth,trim={0cm 0cm 0cm 0.7cm},clip]{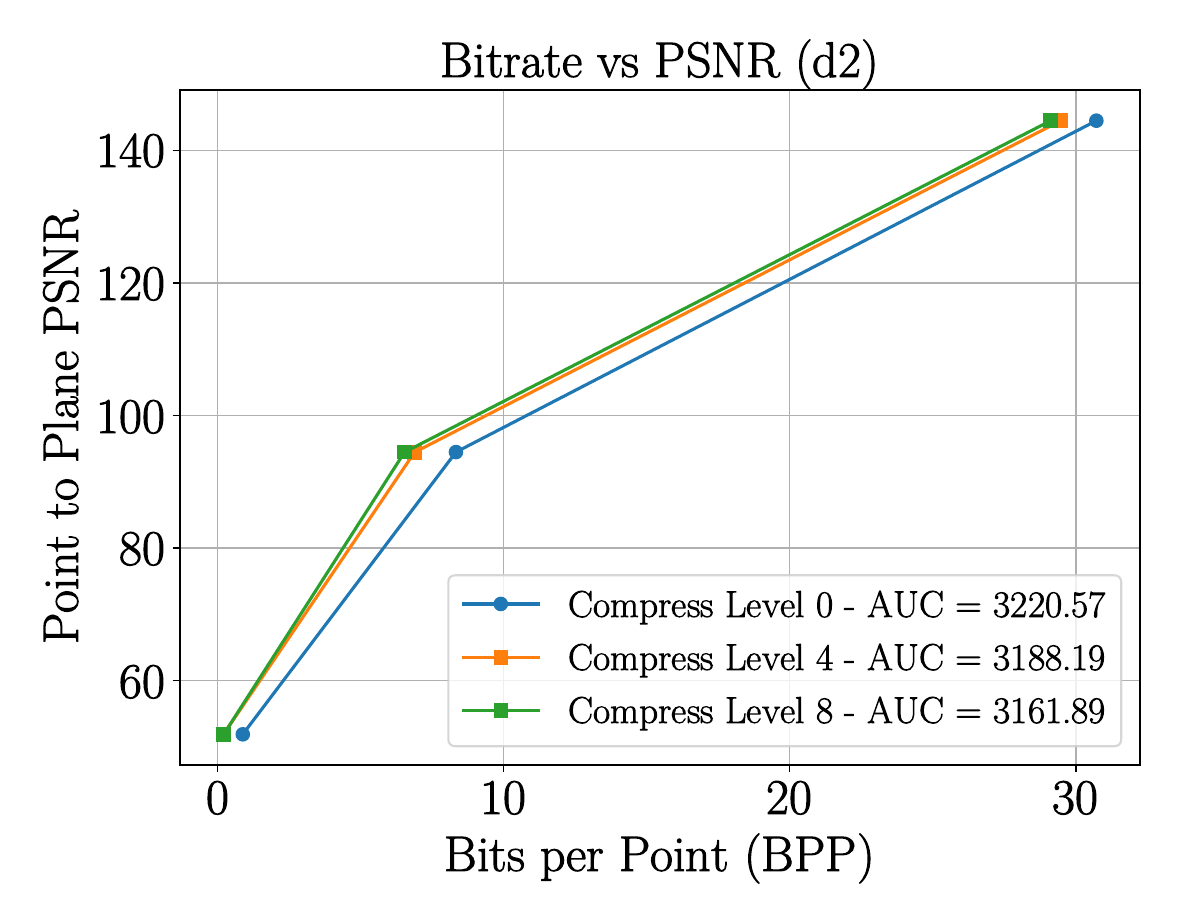}
    \caption{\textbf{Visualization of \textit{BPP} and \textit{PSNR} metrics.} From left to right: \textit{Depoco}, \textit{3DPCC} and \textit{Draco}. We calculate the \textit{Area Under the Curve} (\textit{AUC}) metric for different compression levels and see that \textit{Draco} outperforms the other methods.}
    \label{fig:bpp_vs_psnr}
\end{figure*}

\textbf{Bits per point.} Compression rate is measured using \textit{bits per point} (\textit{BPP}), indicating the bits needed to store a single point. For instance, a binary \texttt{pcd} file with 16-bit precision for $(x,y,z)$ coordinates requires $3\times16 = 48$ bits, which equals 6 bytes. In contrast, \textit{ASCII} format uses 3-36 bytes due to its readable text storage (1 byte per character). Compression formats achieve significantly smaller sizes.

\textbf{Metric Graph.} For each \textit{bits per point} value representing compression method efficiency, there will be a \textit{PSNR} value indicating the reconstruction quality of that specific \textit{BPP} from a certain compression method (see \cref{tab:sota_methods}). We can control \textit{BPP} indirectly in several ways, including quantization level, \textit{octree} depth, or entropy coding efficiency. This will also indirectly affect the \textit{PSNR} value. It is important to note that the \textit{BPP} value is proportional to \textit{PSNR}. When the \textit{BPP} value is higher, \textit{PSNR} will also be higher and vice versa. Therefore, we usually use a 2D graph to map the efficiency and effectiveness of compression methods and compare their performance using those graphs.

\textbf{Compression Speed.} We also measure the compression and decompression speed individually to evaluate the performance of the methods. The speed is calculated only for compression and decompression processes, neglecting the speed required for pre- or postprocessing steps.

\textbf{Memory Consumption.} Finally, we report the allocated memory for the encoding and decoding on the GPU (\textit{VRAM}) for each compression algorithm.

\textbf{Mean Average Precision.} The \textit{mean average precision} (mAP) metric quantifies the average precision scores across multiple object categories, evaluating a detection model's performance across different classes. 
% [] TODO: move this to quantitative results
%For our original uncompressed point cloud data we get an $mAP$ of 20.11.

\subsection{Experiment Setup}
\textbf{Datasets.} We use the \textit{TUMTraf Intersection} \cite{zimmer2023tumtraf} (release R02, sequence S02) and the \textit{TUMTraf V2X Cooperative Perception} \cite{zimmer2024tumtrafv2x} datasets, labeled with 3D BAT \cite{zimmer20193d,zimmer2024tumtrafv2x}, for our experiments. The latter one contains ten 10 s long sequences. Each sequence has 100 LiDAR point cloud scans and their corresponding 2D projection images (range, reflectivity, signal, near-infrared). 

\textbf{Hardware Setup.} Our framework runs inside a docker container under an Ubuntu operating system.
We use 3x NVIDIA GeForce RTX 3090 GPUs for our experiments, each having 24576 MiB of VRAM.

\subsection{Quantitative Results}
In this section, we evaluate different methods on the \textit{TUMTraf datasets} and provide quantitative results for \textit{bits per point}, \textit{PSNR} d1 and d2, encoding and decoding speed.
% We also consider each methods performance towards different additional attributes over the spatial information of the point cloud data.

% \begin{figure}
%     \centering
%     %\captionsetup{type=figure}
%     %\captionof{figure}{Reconstructed Point Cloud Result on different max\_nr\_pts and grid size}
%     % \begin{tabular}{cccc}
%     % 3D $\rightarrow$ 2D $\rightarrow$ 3D &  3D $\rightarrow$ 2D+RG $\rightarrow$ 3D & 3D $\rightarrow$ compress(2D) $\rightarrow$ 3D \\ \includegraphics[align=c,width=0.3\textwidth,frame]{figs/rwth_ori.png} &\includegraphics[align=c,width=0.3\textwidth,frame]{figs/rwth_change_range.png} & \includegraphics[align=c,width=0.3\textwidth,frame]{figs/rwth_compress.png}
%     % \\
%     % PSNR 37.247 & PSNR 41.45 & PSNR 41.45 (db)
%     % \\
%     % \end{tabular}
%     %3D $\rightarrow$ 2D $\rightarrow$ 3D\\
%     %\vspace{1mm}
%     \includegraphics[align=c,width=1.0\linewidth]{figs/3d_2d_3d.pdf}\\
%     %\vspace{1mm}
%     %PSNR 37.247
%     \caption{Qualitative result of the original reconstruction from the 2D images acquired from 3D to 2D transformation.\label{fig:rwth-reconstruction}}
% \end{figure}%

\begin{figure*}[bht]
    \centering
    \includegraphics[width=1\linewidth,frame,clip]{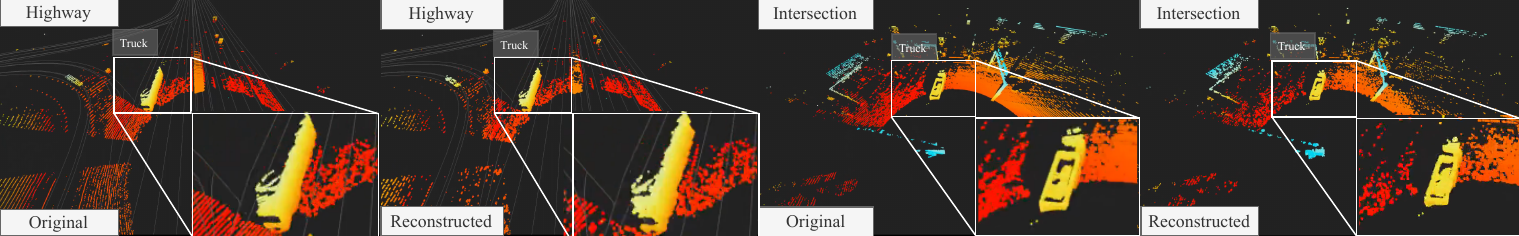}
    \caption{\textbf{Qualitative reconstruction results of \textit{Depoco}}. Left to right: 1) Original and reconstructed point cloud of the TUM Traffic A9 Highway (TUMTraf-A9) dataset \cite{cress2022a9}. 2) Original and reconstructed point cloud of the TUM Traffic Intersection (TUMTraf-I) dataset \cite{zimmer2023tumtraf}.}
    \label{fig:qualitative_results_highway_intersection_depoco}
\end{figure*}

\begin{figure}[ht]
    \centering
    \includegraphics[width=1\linewidth,frame,clip]{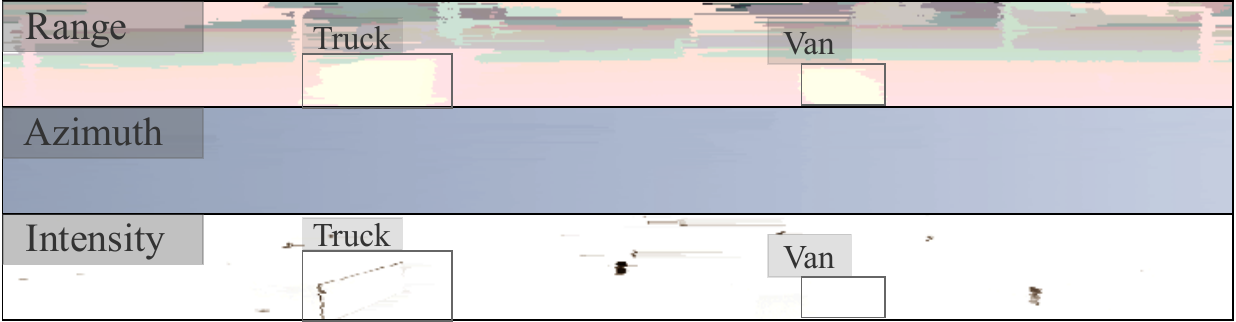}
    \caption{\textbf{Results of the 3D to 2D transformation.} From top to bottom: range, azimuth, and intensity image. We post-process the images and show the front part of the LiDAR.}
    \label{fig:conversion}
\end{figure}

%Referring to Sec. \ref{chap:depoco}, 
We find that the \textit{max. number of points} and \textit{grid size} are the most impactful parameters for Depoco. 
We present the results of our parameter fine-tuning in Table \ref{tab:depoco-tune}. 
The decoding speed remains consistent across configurations, while the encoding speed, GPU memory usage, and \textit{BPP} increase with a higher max. number of points. 
We calculate the \textit{BPP} by dividing the encoded binary file size by the number of points in the original point cloud. 
We proceed by analyzing the conversion of 3D point clouds to 2D images. 
Each 3D point cloud frame yields three images (AIR): azimuth, intensity, and range. 
An example result is shown in Fig. \ref{fig:conversion}. 
The transformation speed from 3D to 2D points is 300 ms, and from 2D to 3D points is 19 ms, while the \textit{ImageMagick} PNG to JPG conversion takes approximately 15 ms. 
The average size of the range, azimuth, and intensity images from conversion compared to the original ones can be seen in \cref{table:image-compression-size}. 
Further, we compress the images using \textit{PNG} compression using the \textit{PIL} python library, online compression \textit{TinyJPG} \cite{tinyjpg}, and ImageMagick library with compression level 100 for the range image only. 
As we can see from the table, the \textit{PNG} compression by \textit{PIL} hardly compresses the images followed by the \textit{ImageMagick}, while the \textit{TinyJPG} performs really well. 
At last, we evaluate the \textit{Draco} method on the TUMTraf-I dataset, achieving encoding and decoding times of 60 ms and 80 ms, a \textit{PSNR} of 52 dB and 144 dB, and an \textit{mAP} of 20.01 and 20.11, respectively. 
Finally, we compare \textit{Depoco's} performance to \textit{3DPCC} and \textit{Draco}, visualizing it in a 2D graph (see Fig. \ref{fig:bpp_vs_psnr}).

% \begin{figure}[h]
%     \centering
%     \includegraphics[width=1\linewidth]{figs/conversion_range.png}\hfill
%     \includegraphics[width=1\linewidth]{figs/conversion_azi.png}
%     \includegraphics[width=1\linewidth]{figs/conversion_inentisy.png}
%     \caption{Result of the 3D to 2D transformation. From top to bottom: range, azimuth, and intensity image.}
%     \label{fig:conversion}
% \end{figure}

    % \begin{table}[htbp]
    %     \centering
    %     \caption{Reconstruction Error on TUMTraf with Varying Bpp values}
    %     \begin{tabular}{p{0.2\linewidth}p{0.2\linewidth}p{0.2\linewidth}p{0.2\linewidth}}
    %     \toprule
    %     \textbf{Methods} & \textbf{Low Bpp} & \textbf{Med bpp} & \textbf{High bpp} \\
    %     \midrule
    %     GPCC & - & - & - \\
    %     Draco & - & - & - \\
    %     2D Projection & - & - & - \\
    %     Depoco & - & - & - \\
    %     \bottomrule
    %     \end{tabular}
    %     \label{table:bpp-sota}
    % \end{table}

\subsection{Qualitative Results}
% [] TODO: this is another LIMITATION. Why only depoco?
In \cref{fig:depoco-params} we show the reconstruction results for different parameters of \textit{Depoco} that were fine-tuned and listed in \cref{tab:depoco-tune}. 
% [] TODO: figure reference missing:
%Furthermore, we provide qualitative results for the reconstruction from range images as depicted in \cref{table:image-compression-error}. 
We illustrate how the \textit{PSNR} value correlated to the grid size and the number of points. 
Increasing the \textit{max. number of points} parameter will add more points to the scene after decoding and lead to a higher \textit{PSNR} value.
Note that we keep the grid size low, either \texttt{8x8x3} or \texttt{16x16x6}, to get a good reconstruction result. 
Otherwise, reconstructed objects are blurry, as we can see in the \texttt{200k|24x24x9} setting, and it will get even more blurry as the grid size gets larger. 
Keeping the grid size small keeps the \textit{PSNR} high, meaning that the reconstructed point cloud looks more similar to the original one. 
Fig. \ref{fig:depoco_3dpcc_draco_on_tumtraf} illustrates the optimal reconstruction quality achieved at minimal bits per pixel for \textit{Depoco}, \textit{3DPCC}, and \textit{Draco}.
Finally, we show qualitative reconstruction results of \textit{Depoco} on the \textit{TUMTraf-A9} and the \textit{TUMTraf-I} dataset (see \cref{fig:qualitative_results_highway_intersection_depoco}).

% In Fig. \ref{fig:rwth-reconstruction}, we can see that the original reconstruction from the 2D images acquired from 3D to 2D transformation performs the best.
%, while the other two, when we replace the range image with our range image, and we compress the image do not reconstruct well.

\section{Conclusion and Future Work}
\textit{Draco} emerges as the most promising method, with encoding and decoding speeds exceeding 10 FPS alongside good \textit{PSNR} and \textit{mAP} values. \textit{Draco's} \textit{PSNR} and \textit{mAP} values consistently correlate, with higher \textit{PSNR} values leading to better \textit{mAP} values. Quantization bits are revealed as the most influential parameters in \textit{Draco}, with values of 16 yielding \textit{mAP} values of 20.01 and 20.11 when set to the maximum of 30 bits. On the other hand, \textit{Depoco} faces speed issues despite configurations surpassing the baseline \textit{mAP} of 20.11. The surpassing result is due to \textit{Depoco's} introduction of additional points, enhancing object recognition but leading to inconsistent \textit{PSNR-mAP} relationships. 
% [] TODO: "without compression"?? without compression the PSNR value should be 0 or not?
% [] TODO: how high is the PSNR value for 3DPCC? on what dataset?
Lastly, \textit{3DPCC} exhibits a good \textit{PSNR d2} value of 36.23 compared to 32.88 for \textit{Depoco} on the \textit{TUMTraf-I} dataset. 
On the other hand, reconstructed point clouds with 3DPCC lead to a very low 3D object detection performance (\textit{mAP} of 7.49).
We found that fixing the transformation to prevent loss of point clouds can improve that.
Future work includes deploying our framework on the live system, and the extension to other LiDAR types and manufacturers.

\section*{Acknowledgment}

This research was supported by the Federal Ministry of Education and Research in Germany within the project $\text{\textit{AUTOtech.agil}}$, Grant Number: 01IS22088U.

\twocolumn[{%
\renewcommand\twocolumn[1][]{#1}%
%\maketitlesupplementary
%\vspace{0.4cm}
\begin{center}
\Large{\textbf{Supplementary Material}}\\[8pt]
\large{\url{https://pointcompress3d.github.io/}}
\end{center}
%\vspace{0.5cm}
\begin{center}
    \centering
    \captionsetup{type=figure}
    \fbox{\includegraphics[width=1.0\textwidth]{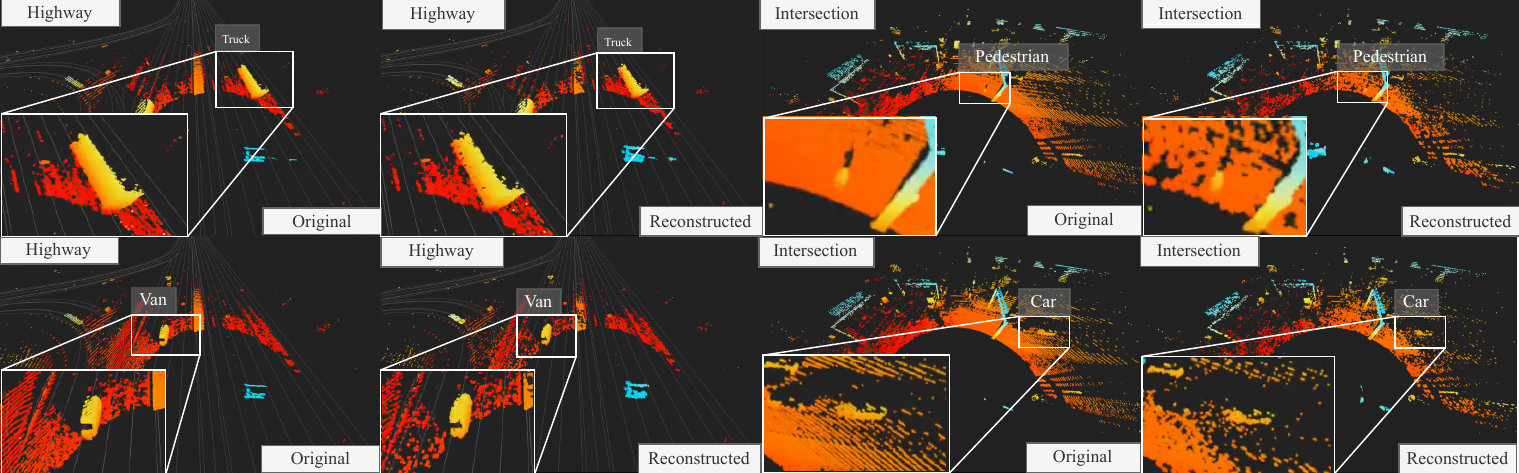}}
    \captionof{figure}{\textbf{Visualization of point cloud reconstruction results.}
  We show further reconstruction results of \textit{Depoco} on the \textit{TUMTraf-A9} and \textit{TUMTraf-I} dataset.
  First row from left to right: 1) Original and reconstructed point cloud on the A9 Highway showing a truck, 2) Original and reconstructed point cloud at an intersection showing a pedestrian.
  Second row from left to right: 3) Original and reconstructed point cloud on the A9 Highway showing a van, 4) Original and reconstructed point cloud at an intersection showing a car.
  }
   % TODO: and 3D object detection results
   % TODO: during day and night time
  \label{fig:qualitative_reconstruction_and_detection_results}
    
\end{center}%
\vspace{0.1cm}
}]

\section*{Contents}
\vspace{-0.3cm}
{
  \hypersetup{linkcolor=black}
  \appendix % all chapter will be automatically in letters inside toc after this command
  \startcontents[sections]
  \printcontents[sections]{l}{1}{\setcounter{tocdepth}{2}}
}
%\clearpage
%%%%%%%%% BODY TEXT
\section{Task definition}
Point cloud compression involves reducing the data size of point clouds, represented as sets of points $\{(x_i,y_i,z_i)\}$ in space, while preserving their geometric and attribute information.
The task aims to minimize the storage $S$ and transmission time $t$ requirements by finding a compact representation $C$ such that $|C| << |P|$, where $P$ denotes the original point cloud data.
The goal of efficient compression algorithms is to achieve high compression rates $r=\frac{|P|}{|C|}$ with minimal loss of information $\epsilon$, thus ensuring that the reconstructed point cloud $P'$ closely approximates $P(d(P,P') \le \epsilon)$.
Compression algorithms are applied in applications like autonomous driving, virtual reality, and geographic information systems, where large volumes of 3D data need to be processed and transmitted efficiently.

\section{Further related work}
% main paper contains 16 compression methods in related work
This section provides an overview of additional point cloud compression methods that have been proposed in the literature.
%    \item \textbf{Point Cloud Compression with Graph Neural Networks} \cite{li2024point}: Proposes a novel point cloud compression method based on graph neural networks that outperforms existing methods in terms of compression rate and reconstruction quality.
%    \item \textbf{Point Cloud Compression with Deep Learning} \cite{zhang2024point}: Introduces a deep learning-based point cloud compression method that achieves state-of-the-art performance on various benchmark datasets.
%    \item \textbf{Point Cloud Compression with Generative Adversarial Networks} \cite{wang2024point}: Presents a generative adversarial network-based point cloud compression method that generates high-quality reconstructions with low bit rates.
%    \item \textbf{Point Cloud Compression with Reinforcement Learning} \cite{liu2024point}: Proposes a reinforcement learning-based point cloud compression method that learns to compress point clouds efficiently while maintaining reconstruction quality.
%    \item \textbf{Point Cloud Compression with Attention Mechanisms} \cite{chen2024point}: Introduces an attention mechanism-based point cloud compression method that focuses on important regions of the point cloud to improve compression efficiency.

%and shows how our proposed method, \textit{PointCompress3D}, compares to existing approaches.
\subsection{Traditional point cloud compression}
Some of the most commonly used traditional point cloud compression algorithms include \textit{MPEG PCC} \cite{schwarz2018emerging} and \textit{AVS PCC} \cite{li2024point}.
\textit{MPEG PCC} is a standardization effort by the \textit{Moving Picture Experts Group} (MPEG) for point cloud compression.
\textit{AVS PCC} is a similar standardization effort by the \textit{Audio Video Coding Standard Workgroup of China} (AVS).
Traditional point cloud compression algorithms are based on geometry coding, attribute coding, and entropy coding techniques.
Yu et al. \cite{yu2023regularized} propose a regularized projection algorithm to construct a reliable prediction relationship in the predictive structure.
A simplified geometry prediction technique is proposed based on the regularized projection pattern.
\cite{ruiu2024saliency} introduces two new encoding methods for point cloud compression, namely \textit{Alternate Depth Compression} (ADC) and \textit{Log-Polar} (LP).
In \cite{luo2024scp} the authors propose \textit{SCP}, a model-agnostic spherical coordinate-based point cloud compression method.
Moreover, a multi-level Octree is introduced to mitigate the reconstruction error for distant areas in the point cloud.

\subsection{Learning based compression}
Learning-based point cloud compression algorithms have shown promising results in terms of compression efficiency and reconstruction quality.
\textit{EHEM} \cite{song2023efficient} proposes a hierarchical attention structure that has a linear complexity to the context scale and maintains the global receptive field.
Furthermore, a grouped context structure is presented to address the serial decoding issue caused by the auto-regression mechanism.
\textit{PIVOT-Net} \cite{pang2024pivotnet} introduces a heterogeneous point cloud compression framework that unifies typical point cloud representations such as point-based, voxel-based, and tree-based representations.
This allows compressing point clouds at different bit-depth levels.
\textit{SparsePCGC} \cite{wang2023sparse} is a sparse point cloud geometry compression method that only performs convolutions on sparsely distributed most-probable positively occupied voxels.

\section{Implementation details}
This section provides additional implementation details of our point cloud compression framework \textit{PointCompress3D}.
%\subsection{Compression architecture}
\subsection{Hardware and software setup}
We run our compression framework on a GPU cluster with an AMD EPYC 7001 series CPU, 128 GB of RAM, and 3 x NVIDIA RTX 3090 GPUs.
The framework runs in a Docker container based on the Ubuntu 20.04 operating system and is implemented in Python.
\subsection{Parameter tuning}
We perform parameter tuning to optimize the performance of our point cloud compression framework.
The parameters include the grid size, the max. number of points, min. kernel size, kernel radius (encoding/decoding), quantization bits and compression level.
We use grid search to find the optimal parameters for each model.
The parameters are tuned on the \textit{TUMTraf-I} dataset and are then used for testing the models on the \textit{TUMTraf-V2X} dataset.

\section{Point Cloud Compression Datasets}
This section provides an overview of the most commonly used point cloud datasets for evaluating point cloud compression algorithms \cite{li2024point}.
\begin{enumerate}
    \item \textbf{KITTI} \cite{geiger2013vision}: Contains point clouds captured by a \textit{Velodyne} LiDAR sensor mounted on a moving vehicle.
    \item \textbf{ShapeNet} \cite{chang2015shapenet}: Consists of 3D models from various categories, including cars, chairs, and tables.
    \item \textbf{ModelNet40} \cite{wu20153d}: Contains 3D models from 40 categories, such as airplanes, chairs, and tables.
    \item \textbf{SemanticKITTI} \cite{behley2019semantickitti}: A large-scale dataset of point clouds with semantic annotations.
    \item \textbf{TUMTraf-I} \cite{zimmer2023tumtraf}: Contains point clouds captured by two Ouster LiDAR sensors mounted on roadside infrastructure.
\end{enumerate}

% MPEG PCC Category 1
% MPEG PCC Category 2
% MPEG PCC Category 3
% AVS PCC Category 1A
% AVS PCC Category 1B
% AVS PCC Category 1C
% AVS PCC Category 2-frame
% AVS PCC Category 3
% MVUB
% KITTI
% ShapeNet
% ModelNet40

\section{Metrics}
The metrics for evaluating point cloud compression performance and reconstruction quality can be categorized into performance metrics and quality metrics \cite{ruiu2024saliency}.
\subsection{Performance metrics}
\begin{enumerate}
\item Bit per point (BPP): Measures the size in bits used to represent a single point.
\item Bit-rate (BR): Total size of the computed frames divided by the frame transmission rate.
\item Compression ratio (CR): Ratio between the \textit{BPP} of the raw point cloud and the \textit{BPP} of the encoded point cloud.
\item Encoding and decoding time (ET, DT): Time taken to process the data and generate the corresponding output.
\end{enumerate}

\subsection{Quality metrics}
\begin{enumerate}
    \item Peak Signal-to-Noise Ratio (PSNR):
        \begin{itemize}
        \item PSNR-D1: Intra-point MSE error calculated between the original and reconstructed point cloud.
        \item PSNR-D2: Similar to PSNR-D1 but measures distances with respect to a plane.
        \end{itemize}
    \item PointSSIM (PSSIM): Measures perceptual degradations of point cloud data using statistical dispersion measurements for attributes such as colors and geometry.
    \item Video Quality Assessment Point Cloud (VQA-PC): Uses spatial and temporal features extracted from a point cloud to estimate quality levels through a machine learning model.
    \item Multi-Modal Point Cloud Quality Assessment (MM-PCQA): The point cloud is split into various 3D sub-models and rendered into 2D image projections.
          These projections are encoded with neural networks, and the quality level is estimated through a quality regression block.
\end{enumerate}
These metrics are used to evaluate both the efficiency of the compression process and the quality of the reconstructed point cloud.

%\subsection{SNNRMSE}
%\subsection{Intersection over union}

%\section{Further experiments and ablations}
%\subsection{Ablation with multiple point clouds}
%\subsection{Ablation with parameter tuning}

\section{Detailed visualization results}
We provide additional qualitative results of the point cloud reconstruction process using \textit{Depoco} on the \textit{TUMTraf-A9} and \textit{TUMTraf-I} datasets.
Figure \ref{fig:qualitative_reconstruction_and_detection_results} shows the original and the reconstructed point cloud on the A9 Highway with a truck and a van.
Moreover, we show the original and reconstructed point cloud at an intersection showing a pedestrian an a car.
These results demonstrate the effectiveness of \textit{Depoco} in reconstructing point clouds with different objects and scenes in the \textit{TUMTraf} datasets.

%\subsection{Depoco}
%\subsection{3DPCC}
%\subsection{Draco}

\section{Failure cases and limitations}
\subsection{Failure cases}
Learning-based point cloud compression algorithms like \textit{3DPCC} \cite{beemelmanns20223d} can encounter several failure cases.
These include overfitting to the training data, resulting in poor performance on unseen data, and underfitting, where the model fails to capture underlying patterns.
They may also struggle with data imbalance, performing poorly when one class is overrepresented, and with outliers, which can negatively impact the model's performance on datasets that contain atypical data points.

\subsection{Limitations}
\textit{Depoco} \cite{wiesmann2021deep} is no longer state-of-the-art (SOTA) and has only been evaluated on the \textit{SemanticKITTI} dataset.
It is outperformed by \textit{Draco} on the \textit{TUMTraf-I} dataset and does not consider the compression of point attributes.
Similarly, \textit{3DPCC} \cite{beemelmanns20223d} also does not consider the compression of attributes except the intensity and XYZ coordinates.
It requires calibration data, range images, and azimuth maps for training, which are not always available in practice.
It is also outperformed by \textit{Draco} \cite{galligan2018draco} on \textit{TUMTraf-I} and is prone to overfitting.
Finally, \textit{Draco} is not as efficient as \textit{3DPCC} or \textit{Depoco} in terms of decoding point clouds.
\pagebreak
%%%%%%%%% REFERENCES
%\addtolength{\textheight}{-4.8cm}
\balance
{\small
\bibliographystyle{ieee_fullname}
\bibliography{references}
}

\end{document}